\documentclass[onecolumn,showpacs,aps,amssymb,floatfix,prd,amsmath,preprintnumbers]{revtex4}
\usepackage{epstopdf}
\usepackage{capt-of}
\usepackage{graphicx}  % Include figure files
\usepackage{dcolumn}   % Align table columns on decimal point
\usepackage{float}
\usepackage{hyperref}
\begin{document}
%\input epsf.tex
%%%%%%%%%%%%
%%%%%%%%%%%
\title{Power-law solution for homogeneous and isotropic universe in $f(R,T)$ gravity}

\author{Lokesh Kumar Sharma$^{1}$\footnote{Email: lksharma177@gmail.com},
Anil Kumar Yadav $^{2}$\footnote{Email: abanilyadav@yahoo.co.in},    
B. K. Singh$^{3}$\footnote{Email:  benoy.singh@gla.ac.in}}\
\affiliation{$^{1,3}$Department of Physics, GLA University, Mathura - 281406 India}
\affiliation{$^{2}$Department of Physics, United College of Engineering and Research, Greater Noida - 201306, India}  

%%%%%%%%%%%%
\begin{abstract}

\textbf{Abstract:} In the present work, we investigate power-law solution for homogeneous and isotropic universe in $f(R,T)$ gravity by considering its functional form $f(R,T) = R + \xi R T$ with $\xi$ being positive constant. We have constructed the field equation in $f(R,T)$ gravity for homogeneous and isotropic space time. The explicit solution of the constructed model is obtained by considering the scale factor as $a  = \alpha t^{\beta}$ with $\alpha$ and $\beta$ being free parameters. The values of $\alpha$ and $\beta$ are obtained by using Markov Chain Monte Carlo (MCMC) method to constraining the model under consideration with observational $H(z)$ data. Some physical and kinematic properties of the model are also discussed.\\



\end{abstract}
\maketitle

\section{Introduction}
The recent astrophysical observations of H(z) data of SN Ia, CMB anisotropy and Plank collaboration have showed strong observational evidence that we are living in accelerating universe at present time \cite{Perlmutter/1998,Riess/2004,Eisenstein/2005,Aghanim/2017}. However, it was also observed that the present universe was in decelerating phase in it's beginning. The discovery of late time acceleration of universe is surprising and posed a major challenge to theoretical Physicist and Cosmologist to know the exact reason of this acceleration \cite{Mishra/2019}. In general theory of relativity, it has been revealed that the accelerated expansion of the present universe is due to some form of non-baryonic energy which have anti gravitation effect and increase the rate of expansion of universe \cite{Peeble/2003}. So, the cosmological constant $\Lambda$ returns in the story with new role and leads the accelerated expansion at present epoch. But yet, it could not well defined with respect to the fine-tuning and cosmic coincidence puzzles\cite{Copeland/2006}. Therefore, Caldwell \cite{Caldwell/2002} have credit to initiate the modeling of accelerating universe by characterizing the dynamical nature of dark energy/exotic energy with effective equation of state parameter $(\omega^{de} = \frac{p^{de}}{\rho^{de}} \neq -1)$. Later on, Copeland et al. \cite{Copeland/2006} have reviewed the observational evidence for late time accelerated expansion of universe and presented scalar field dark energy models such as phantom, quintessence  and tachyon models. Some applications of dynamical dark energy with variable equation of state parameter for spatially homogeneous and an-isotropic space-time are investigated in the references \cite{Kumar/2011a,Yadav/2012a,Mishra/2018,Mishra/2015,Ali/2016,Goswami/2016,Yadav/2016}. Some important applications of $\Lambda$-dominated era are given in Refs. \cite{Bhatti/2016,Yousaf/2017a,Yousaf/2017b}. For example, in Refs. \cite{Bhatti/2016,Yousaf/2017a}, the authors have investigated some exact analytical collapsing solutions under the effect of $\Lambda$ with shear-free condition. Later on, Yousaf \cite{Yousaf/2017b} has studied the effect of $\Lambda$ and expansion-free condition on the formulation of an exact analytical solution of relativistic stellar interior by assuming a non-static and non-diagonal cosmic stellar fillament filled with non-isotropic fluids. Alternatively, in the recent time, the modification in general relativity attracts cosmologists to investigate the exact reason of late time acceleration without cosmological constant or exotic energy and hence led down the cosmological constant problems. In 2011, Harko et al \cite{Harko/2011} have proposed $f(R,T)$ theory of gravitation which leads the modification in geometrical part of the Einstein – Hilbert action. In this theory, the matter Lagrangian is function of Ricci scalar as well as trace T of the energy - momentum tensor. One important notion of $f(R,T)$ gravity is that it consider the quantum effect and leads the possibilities of particle production. Such possibilities are important for astrophysical research because it predicts that there is a bridge between quantum theory and $f(R,T)$ gravity. Some important applications of $f(R,T)$ theory of gravitation in astrophysics as well as in cosmology are read in the references \cite{Zubair/2016,Yousaf/2017,Moraes/2017jcap,Das/2016,Singh/2015ijtp,Myrzakulov/2012,Houndjo/2012,Jamil/2012}. Recently Zaregonbadi et al \cite{Zaregonbadi/2016} have checked the viability of $f(R,T)$ gravity in order to investigate the dark-matter effects on the galaxy scale. The influence of $f(G,T)$ corrections on some dynamical properties of evolving stellar objects and the hydrodynamic properties of dissipative fluids in $f(R)$ geometry are respectively described in Refs. \cite{Yousaf/2019a,Yousaf/2019b}. \\

The standard FRW cosmological models prescribe an isotropic and homogeneous distribution of matter inside it for description of current fate of universe. To follow up the observational results, one is bound to construct of spherically symmetric or flat and spatial homogeneous cosmological model of universe \cite{Yadav/2010}. But in the beginning, close to big bang, the present universe could have not had such a smoothed picture. Thus, the inhomogeneity in space-time is the feature of early universe and it play significant role in structure formation in the early stage of evolution of universe and homogenization. Firstly, Bondi \cite{Bondi/1947} had investigated spherically symmetric but inhomogeneous cosmological model and describes many essential features of universe in it's early stage of evolution. Some pioneer investigations for inhomogeneous space-time had been reported by Taub \cite{Taub/1951,Taub/1956}, Senovilla \cite{Senovilla/1990} and \cite{Bolejko/2011}. Romano \cite{Romano/2012} has analyzed the inhomogeneous cosmological model with $H_{0}$ observations and fit their model with observational data. In the recent years, various inhomogeneous cosmological models have been studied by numerous authors \cite{Ali/2016,Yadav/2018,Ellis/2011,Marra/2011,Anderson/2011,Gurses/2019} in different physical contexts. But after discovery of accelerating universe \cite{Perlmutter/1998,Riess/2004}, the homogeneous cosmological models have gained interest in astrophysical community and are more often employed for study of cosmological features of universe. In this paper, our goal is to investigate the simplest cosmological model in modified gravity which have consistency with observations. The recent astrophysical observations confirm that we are living in homogeneous and isotropic universe. So, we assume a flat, homogeneous and isotropic space-time to derived the various characteristics of present universe. In 1996, Carvalho \cite{Carvalho/1996} has investigated a flat FRW cosmological model to describe the early phases of evolution of universe. Later on Singh et al \cite{Singh/2007} have investigated bulk viscous FRW universe and showed that present acceleration in universe is driven by bulk viscous fluid. In the literature, there are some impressive investigations for constraining the various parameters of homogeneous and isotropic universe in different theories of gravitation \cite{Aditya/2019,Kumar/2019,Kumar/2019a,Akarsu/2018}.\\

Motivated by the above discussion, in this paper, our aim is to investigate cosmological model of accelerating universe without inclusion of cosmological constant/dark energy in the framework of $f(R,T)$ theory of gravitation. The paper is organized as follows: Section II deals with the metric and basic formalism of $f(R,T) = f_{1}(R)+f_{2}(R)f_{3}(T)$ gravity. In section III, the physical and kinematic parameters of the model are given. We devote section IV for constraining the model parameters with observational data. Section V represents the physical properties and dynamical behaviour of model in detail. Finally, we have summarized our findings in section VI.
%%%%%%%%%%%%%%%%%%%%%%%%%%%%%%%%%%%%%%%%%%%%%%%%%%%%
\section{The metric and $f(R,T) = f_{1}(R)+f_{2}(R)f_{3}(T)$ gravity}\label{2}
The FRW space-time is read as
\begin{equation}
\label{ref1}
 ds^{2}=-c^{2}dt^{2}+a^{2}\left(dx^{2}+dy^{2}+dz^{2}\right)
\end{equation}
Here, $a$ is the scale factor and it is function of $t$ only.\\
The action in $f(R,T)$ gravity is given by
\begin{equation}
 \label{basic}
S = \frac{1}{16\pi}\int{d^{4}x\sqrt{-g}f(R,T)} +\int{d^{4}x\sqrt{-g}L_{m}}
\end{equation}
where $g$ and $L_{m}$ are the metric determinant and matter Lagrangian density respectively.\\
The gravitational field of $f(R,T)$ gravity is given by
\[
 [f_{1}^{\prime}(R)+f_{2}^{\prime}(R)f_{3}^{\prime}(T)]R_{ij}-\frac{1}{2}f_{1}^{\prime}(R)g_{ij}+
 \]
 \[
(g_{ij}\nabla^{i}\nabla_{i}-\nabla_{i}\nabla_{j})[f_{1}^{\prime}(R)+f_{2}^{\prime}(R)f_{3}^{\prime}(T)]=
\]
\begin{equation}
 \label{basic1}
[8\pi+f_{2}^{\prime}(R)f_{3}^{\prime}(T)]T_{ij}+f_{2}(R)\left[f_{3}^{\prime}(T)p+\frac{1}{2}f_{3}(T)\right]g_{ij}
\end{equation}
Here, $f(R,T) = f_{1}(R)+f_{2}(R)f_{3}(T)$ and primes denote derivatives with respect to the arrangement.\\
Following Moraes and Sahoo \cite{Moraes/2017}, we assume $f_{1}(R) = f_{2}(R) = R$ and $f_{3}(T) = \xi T$, with
$\xi$ as a constant.\\
Thus the equation (\ref{basic1}) yields
\begin{equation}
 \label{basic2}
G_{ij} = 8\pi T^{eff}_{ij} = 8\pi(T_{ij}+T^{ME}_{ij})
\end{equation}
where, $G_{ij}$, $T^{eff}_{ij}$, $T_{ij}$ and $T^{ME}_{ij}$ represent Einstein curvature tensor, the effective energy momentum tensor, matter energy momentum tensor and extra energy term due to trace of energy-momentum tensor T respectively \cite{Bhardwaj/2019}. 
\begin{equation}
 \label{basic3}
T^{ME}_{ij}=\frac{\xi R}{8\pi}\left(T_{ij}+\frac{3\rho-7p}{2}g_{ij}\right)
\end{equation}
Here, $p$ and $\rho$ are the pressure and energy density of perfect fluid.\\
By applying the Bianchi identities in equation (\ref{basic2}) yields
\begin{equation}
 \label{basic4}
\nabla^{i}T_{ij} = -\frac{\xi R}{8\pi}\left[\nabla^{i}(T_{ij}+pg_{ij})+
\frac{1}{2}g_{ij}\nabla^{i}(\rho-3p)\right]
\end{equation}
%%%%%%%%%%%%%%%%%%%%%%%%%%%%%%%%%%%%%%%%%%%%%%%%%%%%%%%%%%%%%%%%%%
Equation (\ref{basic1}) can be written as the following:

\begin{equation}
 \label{basic5}
G_{ij}=\left(8\,\pi+\xi\right)\,T_{ij}-\left(1+\xi\right)\,R_{ij}+\frac{R}{2}\,\Big[1+\xi\,\left(2\,p+T\right)\Big]\,g_{ij}.
\end{equation}
$$T_{ij}=\,\left(\rho+p\right)\,u_i\,u_j-p\,g_{ij}.$$
Here, we take $c\,=\,1$;
$u^i\,=\,(1,0,0,0)\,$, $\,u_i\,=\,(-1,0,0,0)\,$, $\,g_{ij}\,u^i\,u^j\,=\,-1$,
$$\frac{R}{2}\,=-\,3\,\left(\frac{\dot{a}^2}{a^2}+\frac{\ddot{a}}{a}\right).$$
$T_{00}\,=\,2\,p+\rho\,$, $\,T_{ii}\,=\,-a^2\,p\,$, $\,i=1,2,3$.
$$T\,=\,3\,p+\rho.$$
where $\dot{a}$ denotes the differentiation of $a$ with respect to $t$.\\
Solving equation (\ref{basic5}) with line-element (\ref{ref1}), we obtain the following system of equations
\begin{equation}
\label{basic5-1}
3H^{2} = 8\pi \left[\rho-\frac{3 \xi}{8\pi}(\dot{H}+2H^{2})(3\rho-7p)\right]
\end{equation}
\begin{equation}
 \label{basic5-2}
 2\dot{H}+3H^{2} = -8\pi\left[p+\frac{9\xi}{8\pi}(\dot{H}+2H^{2})(\rho-3p)\right]
\end{equation}
%%%%%%%%%%%
where $H$ is the Hubble's parameter and it is defined as $H = \frac{\dot{a}}{a}$.\\

The equations (\ref{basic5-1}) and (\ref{basic5-2}) are the system of two equations with three unknown variable $H$, $\rho$ and $p$. Therefore one can not solve these equation in general. To get the explicit solution for $\rho$ and $p$, we have to assume a parametrization scheme with the requirement of its theoretical consistency and observational viability. For this sake, we can quote the power law expansion \cite{Kumar/2012}.
%%%%%%%%%%%%%%%%%%%%%%%%%%%%%%%%%%%%%%%%%%%%%%%%%%%%%%%%%%%%%%%%%%%%%%%%%%%
\section{Modeling with power law }
In model (\ref{ref1}), $a(t)$ is an arbitrary function of time. So, one can take $a(t)\,=\,\alpha\,t^{\beta}$ with $\alpha $ and $\beta$ are being constants. This form of $a(t)$ describes the power law cosmology and resembles with the late time acceleration of universe. Some important applications of power law cosmology are given in the References \cite{Kumar/2012,Kumar/2011astr,Yadav/2011,Yadav/2011a,Kumar/2011mpla,Yadav/2011b,Sharma/2018}.\\

The space-time (\ref{ref1}), in this case read as
\begin{equation}
\label{ref1-2}
 ds^{2}\,=\,-dt^{2}+\alpha^{2}t^{2\beta}\,\left(dx^{2}+dy^{2}+dz^{2}\right),
\end{equation}
where $\alpha$ and $\beta$ are arbitrary constants.\\

Following, Feinstein and lbanez \cite{dec1} and Raychaudhuri \cite{rayc1}, the deceleration parameter is given by:
\begin{equation}\label{uu1-12a-1}
  \begin{array}{ll}
q\,=\,-\frac{a{\ddot{a}}}{\dot{a}^{2}} =  -\frac{-1+\beta}{\beta}
  \end{array}
\end{equation}

The Hubble parameter $H$, scalar expansion $\Theta$ and proper volume $V$ are respectively found to have the following expressions~\cite{dec1,rayc1}
\begin{equation}
\label{Hubble}
H=\frac{\dot{a}}{a} =\frac{\beta}{t}
\end{equation}
\begin{equation}  \label{u215-c}
\Theta\,=\,\frac{3\,\beta}{t},
\end{equation}
\begin{equation}  \label{u218}
V\,=\,a^{3} =\; \alpha^{3}t^{3\beta} .
\end{equation}
The shear tensor is
\begin{equation}  \label{u219}
  \begin{array}{ll}
\sigma_{ij}\,=\,u_{(i;j)}+\dot{u}_{(i}\,u_{j)}-\frac{1}{3}\,\Theta\,(g_{ij}+u_i\,u_j),
\end{array}
\end{equation}
and the non-vanishing components of the $\sigma_i^j$ are
\begin{equation}  \label{u220}
\left\{
  \begin{array}{ll}
 \sigma_1^1\,=\,-\frac{2\,\beta}{t},\\
\\
\sigma_2^2\,=\,\sigma_3^3\,=\,\sigma_4^4\,=\,0.
   \end{array}
\right.
\end{equation}
Then the expressions for pressure $p$ and energy density $\rho$ are are given by
\begin{equation}
\label{p}
p = -\frac{27\xi\beta^{2}(2\beta-1)+\beta(3\beta-2)\left[8\pi t^{2}-9\xi\beta(2\beta-1)\right]}{54\xi^{2}\beta^{2}(2\beta-1)^{2}+\left[64\pi^{2}t^{2}-288\pi\xi\beta(2\beta-1)\right]t^{2}}
\end{equation}
\begin{equation}
\label{rho}
\rho = \frac{21\xi\beta^{2}(3\beta-2)(2\beta-1)+3\beta^{2}[8\pi t^{2}-27\xi\beta(2\beta-1)]}{54\xi^{2}\beta^{2}(2\beta-1)^{2}+\left[64\pi^{2}t^{2}-288\pi\xi\beta(2\beta-1)\right]t^{2}}
\end{equation}
It is to be noted that $\xi = 0$, our model recovers the case general relativity (GR). Thus for $\xi = 0$, the expressions of $p$ and $\rho$ are read as
\begin{equation}
\label{p-1}
p=\frac{\beta(2\beta-2)}{8\pi t^{2}}
\end{equation} 
\begin{equation}
\label{rho-1}
\rho = \frac{3\beta^{2}}{8\pi t^{2}}
\end{equation}
%%%%%%%%%%%%%%%%%%%%%%%%%%%%%%%%%%%%%%%%%% 
\section{Observational constraints on model parameters}
In this section, we bound the model parameters $H_{0}$, $\alpha$ and $\beta$ of the model under consideration in view of observational H(z) dataset in the redshift range $0\leq z \leq 1.965$. The observational H(z) datasets are compiled in the references \cite{Ryan/2018,Akarsu/2014}. Also, the scale factor in connection with redshift is given by
\begin{equation}
\label{a-1}
a = \frac{a_{0}}{1+z} = \,\alpha\,t^{\beta}
\end{equation}
where $a_{0}$ denotes the present value of scale factor.\\
%%%%%%%%%%%%%%%%%%%%%%%%%%%%%%%%%%%%%%%%%%%%%%%%%%%%%%%%%%%%%%%%%%%%%%%%%%%%%%%%%%%%%%%%%%
\begin{figure}[h!]
\centering
\includegraphics[width=13cm,height=12cm,angle=0]{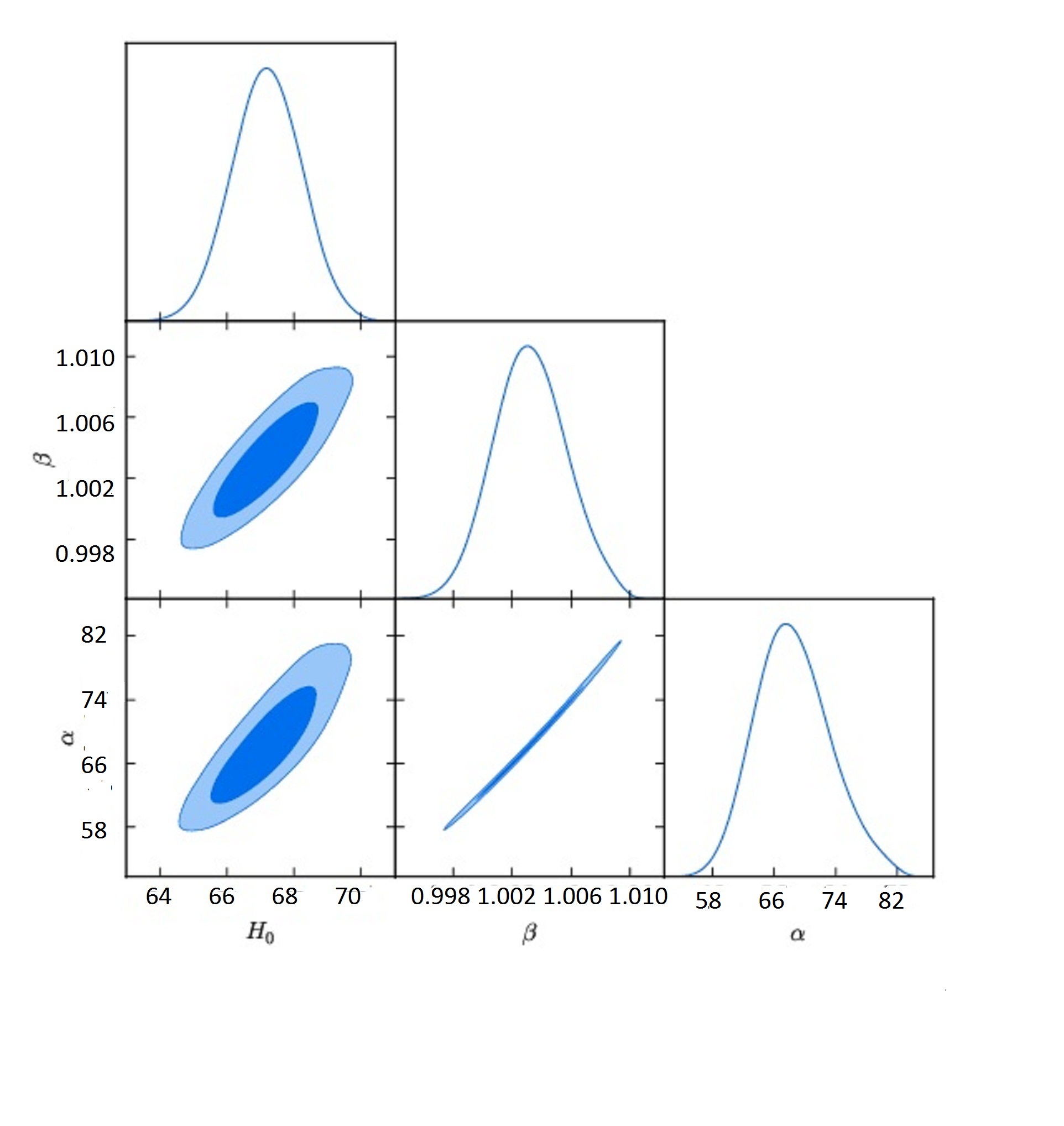}
\caption{One-dimensional marginalized distribution, and two-dimensional contours with $68\%$ confidence level and $95\%$ confidence level.}
\end{figure}
%%%%%%%%%%%%%%%%%%%%%%%%%%%%%%%%%%%%%%%%%%%%%%%%%%%%%%%%%%%%%%%%%%%%%%%%%%%%%%%%%%%%%%%%%%%%%	
%%%%%%%%%%%%%%%%%%%%%%%%%%%%%%%%%%%%%%%%%%%%%%%%%%%%%% Figure 2 %%%%%%%%%%%%%%%%%%%%%%%%%%%
\begin{figure}[h!]
\centering
\includegraphics[width=10cm,height=8cm,angle=0]{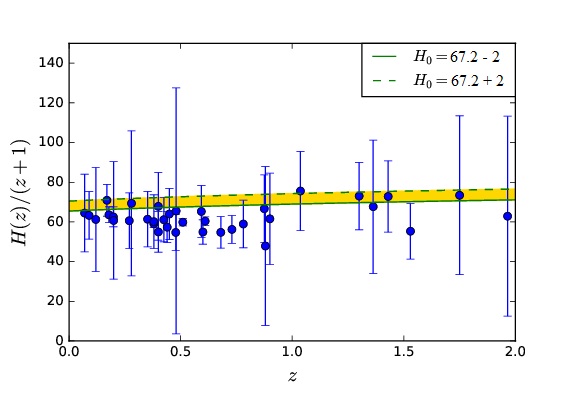}
\caption{The Hubble rate of model for $H_{0}=67.2\pm 2$
		versus the redshift $z$.}
\end{figure}
%%%%%%%%%%%%%%%%%%%%%%%%%%%%%%%%%%%%%%%
The age of universe is computed with following equation
\begin{equation}
\label{age-1}
H(z) = -\frac{1}{1+z}\frac{dz}{dt}
\end{equation}
Taking together equations (\ref{a-1})-(\ref{age-1}) and after some manipulation, one can easily obtain the following expressions for Hubble's function
\begin{equation}
\label{H(z)}
H(z) = \beta\left(\frac{a_{0}}{\alpha}\right)^{-\frac{1}{\beta}}(1+z)^{\frac{1}{\beta}}
\end{equation}
From equation (\ref{H(z)}), the present value of Hubble constant is obtained as $H_{0} = \beta\left(\frac{a_{0}}{\alpha}\right)^{-\frac{1}{\beta}}$.\\
%%%%%%%%%%%%%%%%%%%%%%%%%%%%%%%%%%%%%%%%%%%%%%%%%%%%%%%%%%%%%%%%%%%%%%%%%%%%%%%%%%%%%%%%%%%%%	
\begin{table}[ht]
\centering
{Table~1:~Results from the fits of the model to the H(z) data at 1$\sigma$ and 2$\sigma$ CL}
\vspace{4mm}
\setlength{\tabcolsep}{12pt}
\scalebox{1.12}{
\begin{tabular} {cccc}
\hline
\hline
Model parameters~~&~~ $68\%$~~&~~$95\%$\\[0.2cm]
\hline
\hline
$H_{0}$ ~~~&~~~ $67.2\pm 1.0 $ ~~&~~$67.2^{+2.0}_{-2.0}$ \\[0.2cm]
$\alpha$ ~~~&~~~ $68.5^{+4.3}_{-5.4}$ ~~&~~ $69^{+10}_{-9}$ \\[0.2cm]
$\beta$ ~~~&~~~ $1.0032\pm 0.025 $ ~~&~~ $1.0032^{+0.050}_{-0.048}$\\[0.2cm]
			\hline
\end{tabular}}
\label{table:1}
\end{table}
	
We use observational H(z) data in the redshift range $0\leq z\leq 1.965$ to constrain the model parameters by using Markov chain Monte Carlo (MCMC) method whose code is based on the publicly available package cosmoMC \cite{Lewis/2002}. Also the detail of this method is given in References \cite{Akarsu/2014,Hassan/2018}. Figure 1 depicts the contour plots of model parameters at $68\%$ and $95\%$ confidence level (CL) and robustness of our fits for H(z) of derived model to the data is shown in Figure 2. The summary of numerical analysis is given in table 1. It is worth to note that the estimated value of $H_{0}$ of derived model nicely matches with the result from Plank collaboration \cite{Ade/2016}. In the subsequent sections we check the physical properties  and dynamical behaviour of derived model with $\alpha = 68.5$ and $\beta = 1.032$. These values of $\alpha$ and $\beta$ are obtained by bounding the derived model with observational H(z) data at $68\%$ CL. Note that from equation (\ref{uu1-12a-1}), we observe that the deceleration parameter is negative for $\beta = 1.0032$. We have taken $\alpha = 68.5$ and $\beta = 1.032$ for all the graphical analysis of derived model.\\
%%%%%%%%%%%%%%%%%%%%%%
\section{Physical properties and dynamical behaviour of model}
In this section, we describe the physical properties and dynamical behaviour of the derived model. For this sake, we have explored the importance of dark source terms coming from modified gravity on cosmic evolution. In literature, there are some sensible researches where the usage of dark energy/dark matter corrections have been discussed in the scenario of gravitational collapse \cite{Yousaf/2016prd,Bhatti/2018,Yousaf/2018astr}. In particular, Yousaf et al \cite{Yousaf/2016prd} have described the causes of irregular energy density in $f(R,T)$ gravity. In Bhatti et al \cite{Bhatti/2018}, the authors have checked the viability of wormhole solutions and energy conditions in modified theory of gravity while in Ref. \cite{Yousaf/2018astr}, the role of modification of gravity on some dynamical properties of spherically symmetric relativistic systems is analyzed in detail. Motivated by above researches, in this section, we examine the validation/violation of energy conditions and some physical properties of the derived model in the framework of $f(R,T)$ theory of gravitation.
\subsection{Energy conditions}
%%%%%%%%%%%%%%%%%%%%%   
\begin{figure}[h!]
\centering
\includegraphics[width=9cm,height=7.5cm,angle=0]{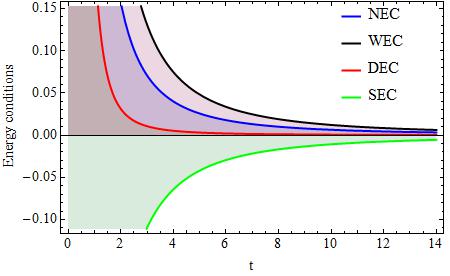}
\caption{Validation/Violation of energy conditions versus time.}
\end{figure}
%%%%%%%%%%%%%%%%%%%%%%%%%%%%%%%
The energy conditions (Ecs) are noted as\\
$$WEC \Leftrightarrow \rho\geq 0$$
$$NEC \Leftrightarrow \rho - p \geq 0$$
$$DEC \Leftrightarrow \rho + p \geq 0$$
$$SEC \Leftrightarrow \rho + 3p \geq 0$$
%%%%%%%%%%%%%%%%%%%%%%%%%%%%%%%%%%
The validation of energy conditions of derived model of universe can be analyzed from equations (\ref{p}) and (\ref{rho})  for $\beta = 1.0032$ and $\alpha = 68.5$. We obtain that except strong Ecs (SEC), all the other energy conditions namely null Ecs (NEC), week Ecs (WEC) and doninant Ecs (DEC) are satisfied. The violation of SEC is in favour of accelerating universe. Thus, the $f(R,T)$ theory of gravity, due to contribution of Trace energy T, is able to give satisfactory description of late time acceleration of present universe without inclusion of cosmological constant or dark energy in the energy content of universe. The graphical representation validation/violation of ECs in our model  is shown in Figure 3.\\
\subsection{Age of universe}
%%%%%%%%%%%%%%%%%%%%%%%
\begin{figure}[h!]
\centering
\includegraphics[width=9cm,height=7.5cm,angle=0]{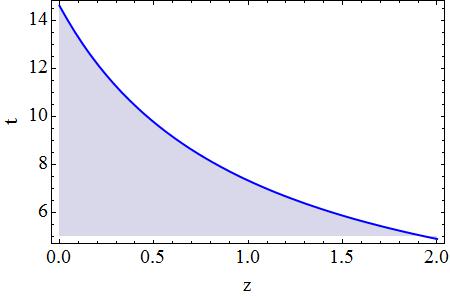}
\caption{Age (in unit of Gyrs) versus redshift $z$.}
\end{figure}
%%%%%%%%%%%%%%%%%%%%%%%%%
The age of derived universe is estimated by integrating equation (\ref{age-1}) as following
\begin{equation}
\label{age-2}
t = -\int \frac{dz}{(1+z)H(z)} = \frac{\beta(z+1)^{-\frac{1}{\beta}}}{H_{0}}
\end{equation}
The present age of universe is obtained by putting $z = 0$ in equation (\ref{age-1}) i. e. $t_{0} = \frac{\beta}{H_{0}}$. Here, we estimate $\beta = 1.0032$ and $H_{0} = 67.2\; km~s^{-1}~Mpc^{-1} \sim 0.0687~Gyrs^{-1}$. Thus the present age of universe is $14.60~Gyrs^{-1}$. The plot of age versus red-shift is shown in Figure 4. \\    
\subsection{Statefinder diagnostic}
To classify the difference among various dark energy cosmological models, firstly Sahni and collaborators \cite{Sahni/2003,Alam/2003} have introduced the pair of statefinder parameter $(r, s)$. Mathematically these parameters are defined as
\begin{equation}
\label{sf}
r = \frac{\ddot{\dot{a}}}{aH^{3}}\;\;\& \;\; s = \frac{(-1+r)}{3(-\frac{1}{2}+q)}
\end{equation}
For Model (\ref{ref1-2}), the expressions for $r$ and $s$ in terms of $q$ are respectively given by
\begin{equation}
\label{r}
r = q(1+2q),\;\;\& \;\; s = \frac{2}{3}(q+1)
\end{equation}
The trajectories of scale factor in derived model have been graphed in Figure 5. Our model follows the result of power law cosmology \cite{Kumar/2012,Sharma/2019a,Rani/2014} for cosmological diagnostic pair. 
\begin{figure}[h!]
\centering
\includegraphics[width= 4.80cm,height=4.0cm,angle=0]{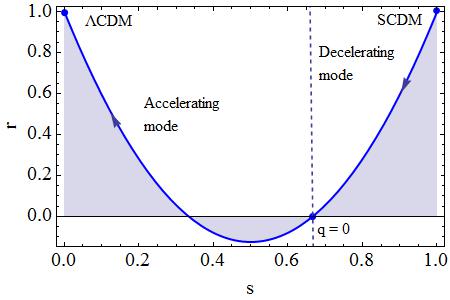}
\includegraphics[width= 4.80cm,height=4.0cm,angle=0]{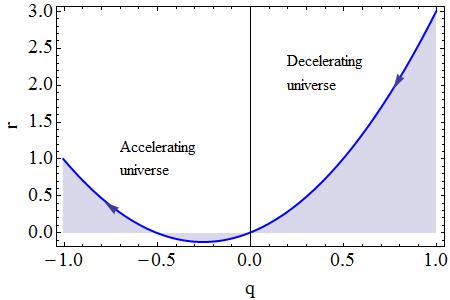}
\includegraphics[width= 4.80cm,height=3.9cm,angle=0]{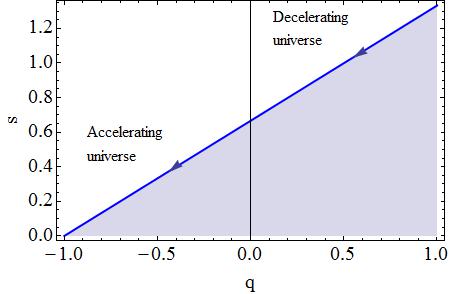}
\caption{Trajectory of scale factor in derived model in $s - r$ plane (left panel), $q - r$ plane (middle panel) and $q - s$ panel (right panel). SCDM and $\Lambda$CDM denote the standard cold dark matter universe $(\Lambda = 0)$ and cosmological constant cold dark matter universe $(\Lambda > 0)$ respectively.}
\end{figure} 
%%%%%%%%%%%%%%%%%%
\subsection{Jerk parameter}
\begin{figure}[h!]
\centering
\includegraphics[width=9.0cm,height=7.50cm,angle=0]{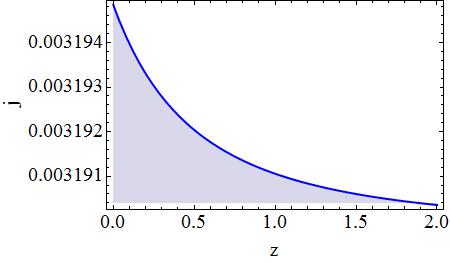}
\caption{Dynamics of jerk versus z for $\beta = 1.0032$}
\end{figure}     
The jerk parameter (j) for model (\ref{ref1-2}), is given by \cite{Visser/2005,Visser/2004}
\begin{equation}
\label{jerk}
j = 1-(1+z)\frac{H^{'}(z)}{H(z)}+\frac{1}{2}(1+z)^{2}\frac{[H^{''}(z)]^{2}}{H^{2}(z)}
\end{equation}
where $H^{'}(z)$ and $H^{''}(z)$ represents the first and second order derivative of $H(z)$ with respect to z respectively.\\

Equations (\ref{H(z)}) and (\ref{jerk}) lead to the following expressions for $j$
\begin{equation}
\label{jerk-1}
j= -\frac{1}{\beta }+\frac{(\beta -1)^2}{2 \beta ^4 (z+1)^2}+1
\end{equation}
%%%%%%%%%%%%%%%%%%%%%%%
The jerk parameter is extensively kinematical quantity which measures more accurately the expansion rate of universe rather than Hubble parameter due to have third order derivative of scale factor with respect to time. A positive jerk parameter expands the universe with acceleration \cite{Nagpal/2019,Rapetti/2007}. From  Figure 6, we see that the derived model evolves with positive values of jerk parameter and it is other than 1. For $\Lambda$CDM dark energy model, $q = -1$ which gives $j =1$. In the model under consideration, $j \neq 1$ but $q$ is negative. Therefore, we can expect other cosmological model of accelerating universe in place of $\Lambda$CDM.\\
%%%%%%%%%%%%%%%%%%%%%%%
\subsection{Particle horizon}
The particle horizon measures the size of observable universe \cite{Bentabol/2013}. The particle horizon is defined as
\begin{equation}
\label{PH}
R_{p} = lim_{t_{p}\rightarrow 0}\;\;a_{0}\int_{t_{p}}^{t_{0}}\frac{dt}{a(t)} = lim_{z\rightarrow \infty}\int_{0}^{z}\frac{dz}{H(z)}
\end{equation}
%%%%%%%%%%%%%%%%%%%
where $t_{p}$ represents time in past at which at which the light signal is transmitted from source.
\begin{figure}[h!]
\centering
\includegraphics[width=9.0cm,height=7.5cm,angle=0]{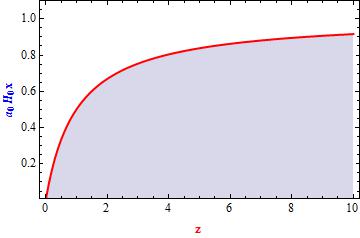}
\caption{Proper distance versus z for $\beta = 1.0032$}
\end{figure} 
%%%%%%%%%%%%%%%%%%%%%%%%%%%%%%%%%%
Integrating equation (\ref{PH}) for large value of red-shift and $\beta = 1.0032$, we obtain particle horizon as $R_{p} = 1.0374\;H_{0}^{-1}$. The dynamics of proper distance versus red-shift is shown in Fig. 7. From Fig. 7, we observe that at present $(i.\; e.\; z = 0)$, $a_{0}H_{0}x$ is null which turn into imply that at $z = 0$, the proper distance $x$ becomes infinite. Thus we are at very large distance ($\sim$ at infinite distance) from an event occurred in the beginning of the universe.
%%%%%%%%%%%%%%%%%%%%%%%%%%%%%  
\section{Conclusion}
In this paper, we have investigated the simplest model of accelerating universe within the framework of $f(R, T)$ theory of gravity by taking its functional form $f(R,T) = R +  \xi RT$. We note that for $\xi = 0$, the derived model recovers the case of general relativity and satisfies all energy conditions for $\alpha = 68.5$ and $\beta = 1.0032$. These values of $\alpha$ and $\beta$ are obtained by constraining the free parameters of derived model with observational data sets. The strong energy condition (SEC) must be violated in accelerating universe that is why we assumed here $f(R,T)$ gravity. It is worth to note that the derived model describes the observable features of present universe including acceleration without inclusion of dark energy or cosmological constant. Some important characteristics of the derived model are as follow:\\
\begin{itemize}
\item[i)] The derived universe has point type singularity at $t = 0$. The energy density and Hubble's parameter diverse at t = 0 which shows that the universe begun with big bang.
\item[ii)] We estimate the present age of universe as 14.60 Gyrs which has pretty consistency with recent observation of Plank collaboration.
\item[iii)] The derived model violates the SEC which in terms imply that $\rho + 3p < 0$.
\item[iv)] The deceleration parameter evolves with negative sign while jerk parameter is found to be positive throughout expansion of the universe. This confirms the late time acceleration of present universe.
\item[v)] In the derived model particle horizon exists which confirms that presently we are at very large distance from an event occurs in the beginning. This behaviour is clearly depicted in Fig. 7.
\item[vi)] The values of $\alpha$ and $\beta$ are estimated by bounding the model under consideration with observational data.
\end{itemize}

As final concluding remarks, we can say that $f(R,T)$ gravity is capable to describing a suitable cosmological model with late time acceleration without contribution of dark energy/cosmological constant. So, one can argue that $f(R,T) = R + \xi R T$ is an alternative of dark energy models and it play non-avoidable role in describing the evolution process of our universe.\\        
\begin{acknowledgements}
The authors wish to express their sincere thanks to the
honourable referee whose valuable comments and suggestions helped in improving
the quality of this paper. Also we are very grateful to 
H. Amirhashchi for his kind assistance in improving some 
technical aspects of the manuscript. 
\end{acknowledgements}


\begin{thebibliography}{99}
\bibitem{Perlmutter/1998} Perlmutter, S. et al., 1998. Nature \textbf{391}, 51.  

\bibitem{Riess/2004} Riess, A. G., 2004. Astron. J. \textbf{607}, 665. 

\bibitem{Eisenstein/2005} Eisenstein, D. J. et al., [SDSS Collaboration], 2005. Astron. J. \textbf{633}, 560.  

\bibitem{Aghanim/2017} Aghanim, N. et al., [Plank Collaboration], 2017. Astron. Astrophys. \textbf{607}, A95.  

\bibitem{Mishra/2019} Mishra, B., Ray, P. P., Myrzakulov, R., 2019. Eur. Phys. J. C \textbf{79}, 34.  

\bibitem{Peeble/2003} Peebles, P. J. E., Ratra, B., 2003. Rev. Mod. Phys. \textbf{75}, 559. 

\bibitem{Copeland/2006} Copeland, E. J., Sami, M, Tsujikava, S., 2006. Int. J. Mod. Phys. D \textbf{15}, 1753. 

\bibitem{Caldwell/2002} Caldwell, R. R., 2002. Phys. Lett. B  \textbf{545}, 23. 

\bibitem{Kumar/2011a} Kumar, S., Singh, C. P., 2011. Gen. Relativ. Grav. \textbf{43}, 1427.

\bibitem{Yadav/2012a} Yadav, A. K., Saha, B., 2012.  Astrophys. Space Sc. \textbf{337}, 759.

\bibitem{Mishra/2018} Mishra, B. et al., 2018.  Astrophys. Space Sc. \textbf{363}, 86. 

\bibitem{Mishra/2015} Mishra, B., Tripathy, S. K., 2015. Mod. Phys. Lett. A \textbf{30}, 1550175. 

\bibitem{Ali/2016} Ali, A. T., Yadav, A. K., Alzahrani, A. K., 2016. Eur. Phys. J. Plus \textbf{131}, 415.

\bibitem{Goswami/2016} Goswami, G. K., Yadav, A. K., Dewangan, R. N., 2016 Int. J. Theor. Phys. \textbf{55}, 4651. 

\bibitem{Yadav/2016} Yadav, A. K., 2016. Astrophys Space Sci \textbf{361}, 276. 

\bibitem{Bhatti/2016} Bhatti, M. Z., 2016. Eur. Phys. J. Plus \textbf{131}, 428

\bibitem{Yousaf/2017a} Yousaf, Z. 2017. Eur. Phys. J. Plus \textbf{132}, 71

\bibitem{Yousaf/2017b} Yousaf, Z. 2017. Eur. Phys. J. Plus \textbf{132}, 276

\bibitem{Harko/2011} Harko, T., Lobo, F. S. N., Nojiri, S., Odintsov, S. D., 2011. Phys. Rev. D. \textbf{84}, 024020. 

\bibitem{Zubair/2016} Zubair, M., Waheed, S., Ahmad, Y., 2016. Eur. Phys. J. C. \textbf{76}, 444. 

\bibitem{Yousaf/2017} Yousaf, Z., Ilyas, M., Bhatti, M. Z., 2017. Mod. Phys. Lett. A. \textbf{32}, 1750163  

\bibitem{Moraes/2017jcap} Moraes, P. H. R. S., Correa, R. A. C., Lobato, R. V., 2017. JCAP \textbf{07}, 029  

\bibitem{Das/2016} Das, A., Rahaman, F., Guha, B. K., Ray, S., 2016. Eur. Phys. J. C. \textbf{76}, 654.  

\bibitem{Singh/2015ijtp} Singh, V., Singh, C. P., 2015. Int. J. Theor. Phys. \textbf{55}, 1257. 

\bibitem{Myrzakulov/2012} Myrzakulov, R., 2012. Eur. Phys. J. C. \textbf{72}, 2203.

\bibitem{Houndjo/2012} Houndjo, M. J. S., 2012. Int. J. Mod. Phys. D. \textbf{21}, 1250003. 

\bibitem{Jamil/2012} Jamil, M., Momeni, D., Raza, M., Mryzakulov, R., 2012. Eur. Phys. J. C. \textbf{72}, 1999. 

\bibitem{Zaregonbadi/2016} Zaregonbadi, R., et al., 2016. Phys. Rev. D. \textbf{94}, 084052. 

\bibitem{Yousaf/2019a} Yousaf, Z. 2019. Eur. Phys. J. Plus \textbf{134}, 245

\bibitem{Yousaf/2019b} Yousaf, Z. 2019. Mod. Phys. Lett. A \textbf{34}, 1950333

\bibitem{Yadav/2010} Yadav, A. K., 2010. Int. J. Theor. Phys. \textbf{49}, 1140. 

\bibitem{Bondi/1947} Bondi, H., 1947. Mon. Not. R. Astro. Soc. \textbf{107}, 410.

\bibitem{Taub/1951} Taub, A. H., 1951. Ann. Math. \textbf{53}, 472. 

\bibitem{Taub/1956} Taub, A. H., 1956. Phys. Rev. \textbf{103}, 454. 

\bibitem{Senovilla/1990} Senovilla, J. M. M., 1990. Phys. Rev. Lett. \textbf{64}, 2219. 

\bibitem{Bolejko/2011} Bolejko, K., Célérier, M. N., Krasiński, A., 2011. Class. Quant. Grav. \textbf{28}, 164002. 

\bibitem{Romano/2012} Romano, A. E., 2012. arXiv:1112.1777v4 [astro-ph.CO] 

\bibitem{Yadav/2018} Yadav, A. K., Ali, A. T., 2018. Int. J. Geom. Methods Mod. Phys. \textbf{15}, 1850026.   

\bibitem{Ellis/2011} Ellis, G. F. R., 2011. Class. Quantum Grav. \textbf{28}, 164001. 

\bibitem{Marra/2011} Marra, V., Notari, A., 2011. Class. Quantum Grav. \textbf{28}, 164004. 

\bibitem{Anderson/2011} Anderson, L., Coley, A., 2011. Class. Quantum Grav. \textbf{28}, 160301. 

\bibitem{Gurses/2019} Gürses, M., Heydarzade, Y., 2019. arXiv:1905.04133 [gr-qc]

\bibitem{Carvalho/1996} Carvalho, J. C., 1996. Int. J. Theor. Phys. \textbf{35}, 2019.

\bibitem{Singh/2007} Singh, C. P., Kumar, S., Pradhan, A., 2007. Class. Quantum Grav. \textbf{24}, 455.

\bibitem{Aditya/2019} Aditya, Y. et al., 2019. Euro. Phys. J. C. \textbf{79}, 1020.

\bibitem{Kumar/2019} Kumar, S., Nunes, R. C., Yadav S. K., 2019. Mon. Not. R. Astron. Soc. \textbf{490}, 1406. 

\bibitem{Kumar/2019a} Kumar, S., Nunes, R. C., Yadav S. K., 2019. Eur. Phys. J. C \textbf{79}, 576.

\bibitem{Akarsu/2018} Akarsu, O., Katirci, N., Kumar, S., 2018. Phys. Rev. D \textbf{97}, 024011.

\bibitem{Moraes/2017} Moraes, P. H. R. S., Sahoo, P. K., 2017. Eur. Phys. J. C \textbf{77}, 480.

\bibitem{Bhardwaj/2019} Bhardwaj, V. K., Rana, M. K., Yadav, A. K., 2019. Astrophys. Space Sc. \textbf{364}, 136. 

\bibitem{Kumar/2012} Kumar, S., 2012. Mon. Not. R. Astron. Soc. \textbf{422}, 2532. 

\bibitem{Kumar/2011astr} Kumar, S., 2011. Astrophys. Space Sc. \textbf{332}, 449. 

\bibitem{Yadav/2011} Yadav, A. K., Yadav, L., 2011. Int. J. Theor. Phys. \textbf{50}, 871.

\bibitem{Yadav/2011a} Yadav, A. K., 2011. Astrophys. Space Sc. \textbf{335}, 565.

\bibitem{Kumar/2011mpla} Kumar, S., Yadav, A. K., 2011. Mod. Phys. Lett. A \textbf{26}, 647. 

\bibitem{Yadav/2011b} Yadav, A. K., Rahaman, F., Ray, S., 2011. Int. J. Theor. Phys. \textbf{50}, 218. 

\bibitem{Sharma/2018} Sharma, L. K., Yadav, A. K., Sahoo, P. K., Singh, B. K., 2018. Results in Physics {\bf 10}, 738.

\bibitem{dec1} Feinstein, A., lbanez, J., 1993. Class. Quantum Grav. \textbf{10}, L227. 

\bibitem{rayc1} Raychaudhuri, A. K., 1979. Theoritical Cosmology, Oxford University Press (First Edition).  

\bibitem{Ryan/2018} Ryan, J., Doshi, S., Ratra, B., 2018. MNRAS \textbf{480}, 759.

\bibitem{Akarsu/2014} Akarsu, O., et al., 2014. JCAP \textbf{01}, 022.

\bibitem{Lewis/2002} Lewis, A., Bridle, S., 2002. Phys. Rev. D \textbf{66}, 103511. 

\bibitem{Hassan/2018} Amirhashchi, H., Amirhashchi, S., 2019. arXiv:1811.05400v4 [astro-ph.CO]  

\bibitem{Ade/2016} Adel, P. A. R. et al., 2016. A \& A \textbf{594}, A13.

\bibitem{Yousaf/2016prd} Yousaf, Z., Bamba, K., Bhatti, M. Z., 2016. Phys. Rev. D \textbf{93}, 124048

\bibitem{Bhatti/2018} Bhatti, M. Z., Yousaf, Z., Ilyas, M., 2018. J. Astrophys. Astron. \textbf{39}, 69.

\bibitem{Yousaf/2018astr} Yousaf, Z., 2018. Astrophys. Space Sc. \textbf{363}, 226.  

\bibitem{Sahni/2003} Sahni, V., et al., 2003. JETP Lett. \textbf{77}, 201. 

\bibitem{Alam/2003} Alam, U., et al., 2003. Mon. Not. R. Astron. Soc \textbf{344}, 10571074. 

\bibitem{Sharma/2019a} Sharma, L. K., Singh, B. K., Yadav, A. K., 2019. arXiv: 1907.03552 [physics.gen-ph]. 

\bibitem{Rani/2014} Rani, S., et al., 2015. JCAP \textbf{2015}, 031. 

\bibitem{Visser/2005} Visser, M., 2005. Gen. Rel. Grav \textbf{37}, 1541. 

\bibitem{Visser/2004} Visser, M., 2004. Class. Quant. Grav \textbf{21}, 2603. 

\bibitem{Nagpal/2019} Nagpal, R., Singh, J. K., Beesham, A., Shabani, H., 2019. arXiv: 1903.08562 [gr-qc] 

\bibitem{Rapetti/2007} Rapetti, D., Allen, S. W., Amin, M. A., Blandford, R. D., 2007. Mon. Not. R. Astron. Soc. \textbf{375}, 1510.

\bibitem{Bentabol/2013} Bentabol, B. M., Bentabol, J. M., Cepa, J., 2013. J. Cosmol. Astropart. Phys. \textbf{02}, 015. 

\end{thebibliography}

\end{document}